\documentclass[ ] {aipproc}
\layoutstyle{6x9}
                                                                                                                     
\begin{document}

\title{QPO Evolution in 2005 Outburst of the Galactic Nano Quasar GRO J1655-40}

\classification{}
\keywords      {Black Holes, X-ray sources, Shock waves}

\author{D. Debnath}{
  address={Indian Centre For Space Physics,
        43 Chalantika,Garia Station Road, Kolkata 700084, India}
}

\author{A. Nandi}{
  address={Indian Centre For Space Physics,
        43 Chalantika,Garia Station Road, Kolkata 700084, India}
  ,altaddress={On deputation from ISRO-HQ, New BEL Road, Bangalore, 560094}
}

\author{P.S. Pal}{
  address={Indian Centre For Space Physics,
        43 Chalantika,Garia Station Road, Kolkata 700084, India}
}

\author{S.K. Chakrabarti}{
  address={S.N. Bose National Center for Basic Sciences,
       JD-Block, Salt Lake, Kolkata,700098, India}
  ,altaddress={Indian Centre For Space Physics,
        43 Chalantika,Garia Station Road, Kolkata 700084, India}
}

\begin{abstract}
GRO J1655-40 showed significant X-ray activity in the last week of
February, 2005 and remained active for the next 260 days. The rising and 
the decline phases of this particular outburst show evidence for systematic movements 
of the  Comptonizing region, assumed to be a CENBOL, which causes the Quasi-periodic 
Oscillations or QPOs. We present both the spectral and the timing results of the 
RXTE/PCA data taken from these two hard spectral states. Assuming that the QPOs 
originate from an oscillating shock CENBOL, we show how the shock slowly moves 
in through the accretion flow during the rising phase at a constant velocity and 
accelerate away outward during the later part of the decline phase. By fitting 
the observed frequencies with our solution, we extract time variation of various 
disk parameters such as the shock locations, velocity etc.
\end{abstract}

\maketitle


\section{Introduction}

The galactic nano-quasar GRO J1655-40 is an interesting Low Mass X-ray Binary (LMXB) 
with a primary mass $M = 7.02\pm0.22~M_\odot$ (\cite{Orosz97}, \cite{Vander98}) 
and the companion star mass = $2.3~M_\odot$ (\cite{Bailyn95}) 
located at a distance of $D = 3.2 \pm 0.2$~kpc (\cite{Hje95}) 
The disk has an approximate inclination angle of $\theta = 69.5^\circ\pm0.1^\circ$
(\cite{Orosz97}) with the line of sight. It was first observed by BATSE on board
CGRO on 27th July 1994 (\cite{Zhang94}) 
This source was dormant for $\sim 8$ years since 1997. During the 1996-1997 outburst, 
it exhibited QPOs in $0.1-300$Hz frequency range. Later, \cite{Strohmayer01} 
reported a QPO at $450$Hz. In the last week of February, 2005 it became 
X-ray active (see, \cite{ch05} \& \cite{ch08} and references therein) 
and remained so for the next 260 days before returning to the hard state again. 
It exhibited a very complex spectral behavior during this time, details of which will 
be dealt with elsewhere \cite{Debnath08}.

In this article, we present  a through analysis of the first two weeks of the very 
initial stage (onset phase) and the very last three weeks of the final stage
(decline phase) of the 2005 outburst. We study in detail the behavior
of QPO frequency in these two phases and show that a satisfactory explanation of this
behavior can be obtained if we assume that an oscillating shock which is sweeping through the
disk inward in the rising phase and outward in the decline phase is responsible for the QPOs.

\section{Origin of QPOs in black hole candidates}

Observations of Quasi-Periodic Oscillations (QPOs) in black hole candidates have been reported 
quite extensively in the literature (e.g., \cite{CM00}; 
\cite{Rod04}) 
and they are variously interpreted to be due to trapped oscillations 
and disko-seismology (\cite{Kato00}; \cite{Rod02}), 
oscillations of warped disks (\cite{Shi02}), 
accretion-ejection instability at the inner radius of the
Keplerian disk (\cite{Rod02}, \cite{Var03}), 
global disk oscillations (\cite{Tit00}), 
shock oscillations (\cite{MSC96}; 
\cite{CM00}) etc. The numerical simulations of accretion flows having QPOs have been reported also 
(\cite{MSC96}; \cite{CAM04}; \cite{RCM97}) .
Moreover, \cite{CAM04} showed that the power-density spectra (PDS) of the simulated 
light curves are similar to what are observed. Shock locations were found to 
be a function of the cooling rate (\cite{MSC96}) and they were found to propagate when viscous 
effects are turned on (\cite{CM95}). 
Presently, we will consider shock oscillation solution because, as explained in \cite{MSC96} and below, 
it is conceptually simpler the unknown `blobs' and moreover, shocks are naturally produced in 
sub-Keplerian flows around black holes (See Chakrabarti, this volume). Perturbations inside a 
Keplerian disk has been used (e.g., \cite{TCG99}) 
but it is not clear how these perturbations would survive for long.

In the shock oscillation solution (\cite{MSC96}; \cite{CM00}; \cite{CAM04}), the shock were found 
to oscillate typically at a frequency inverse of the in-fall time in the post-shock region
$t_{infall}\sim  r_s/v \sim  R r_s(r_s-1)^{1/2}$ (\cite{CM00}; \cite{ch05} \& \cite{ch08}). 
Here, $R$ is the shock strength (ratio of the post-shock to pre-shock density), $r_s$ is the 
shock location in units of the Schwarzschild radii $r_g$, and $v$ is the flow velocity 
in the post-shock region $v \sim 1/R (r_s-1)^{-1/2}$ in units of the velocity of light. 
Thus, the instantaneous QPO frequency $\nu_{QPO}$ (in $s^{-1}$) is expected to be,
$$
\nu_{QPO} = t_{s0}/t_{infall}= t_{s0}/[R r_s (r_s-1)^{1/2}].
\eqno{(1)}
$$
Here, $t_{s0}= c/r_g=c^3/2GM$ is the inverse of the light crossing time of the black hole
of mass $M$ in $s^{-1}$ and $c$ is the velocity of light. In a drifting shock scenario,
$r_s=r_s(t)$ is the time-dependent shock location given by,
$$
r_s(t)=r_{s0} \pm v_0 t/r_g.
\eqno{(2)}
$$
Here, $r_{s0}$ is the shock location when $t$ is zero and $v_0$ is the shock velocity
in c.g.s. units. The positive sign in the second term is to be used for an outgoing
shock in the decline phase and the negative sign is to be used for the in-falling shock
in the onset phase. Here, $t$ is measured in seconds since the first detection of the QPO.

The physical reason for the oscillation of shocks is straightforward. It due to the 
resonance between the cooling time scale in the post-shock region and the 
infall time scale. This oscillation can also take place when the Rankine-Hugoniot 
shock conditions are not fulfilled even though two saddle type sonic points are present 
and therefore no steady shock wave is possible. Thus, the QPO frequency directly gives 
an estimate of the shock location (Eq. 1). In many objects cases, QPO frequencies are 
khown to rise with luminosity (\cite{ST06}). 
In our solution, since an enhancement of the accretion rate increases the local density 
and thus the cooling rate, the resulting drop of the post-shock pressure reduces 
the shock location and increases the oscillation frequency. Also, \cite{CM00} and \cite{Rao00} 
showed that QPOs from the Comptonized photons have the tendency to have a higher $Q$ value. 
This is consistent with the fact that the post-shock regions are also
the regions of Comptonization of the soft photons \cite{CT95} 
in this TCAF model.

\begin {figure}
\centering
 \includegraphics[width=.55\linewidth]{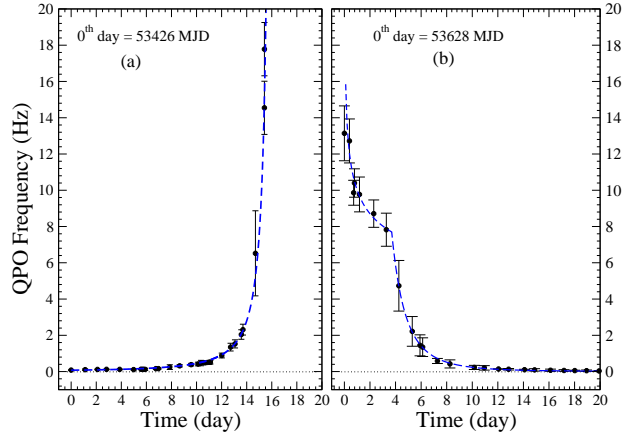}
\caption{Fig. 1 (a-b): Variation of QPO frequency with time (in day)
(a) of the rising phase since the beginning of the outburst and (b) since the
beginning of the declining phase. Error bars are FWHM of fitted Lorenzian curves in the
power density spectrum. The dotted curves are the solutions from oscillating
and propagating shocks. While in (a), the shock appears to be drifting at a
constant speed towards the black hole, in (b) the shock initially moves very
slowly and then runs away at a roughly constant acceleration. According to the
fitted solution, the shock wave goes behind the horizon on the $16.14$th day, about
$15$ hours after the last observed QPO.
}
\end{figure}

\section {Observational results and the analysis}

We concentrate on the data of $51$ Observational IDs (corresponding to a total of $36$ days
of observations) of GRO J1655-40 acquired with the RXTE Proportional Counter Array
(PCA; \cite{Jahoda96}). 
Out of these IDs, $27$ are of onset phase (from MJD 53426 to MJD 53441) and 
$24$ are of decline phase (from MJD 53628 to MJD 53648). We extracted the light curves
(LC), the PDS and the energy spectra from the good detector unit PCU2 which also 
happens to be the best-calibrated. We used FTOOLS software package Version 6.1.1 
and the XSPEC version 12.3.0. For the timing analysis (LC \& PDS), we used the 
Science Data of the Normal mode ($B\_8ms\_16A\_0\_35\_H$) and the Event mode 
($E\_125us\_64M\_0\_1s$, $E\_62us\_32M\_36\_1s$).
For the energy spectral analysis, the {\bf ``Standard2f"} Science Data of PCA was used. 
For all the spectral analysis, we kept the hydrogen column density ($N_{H}$) fixed at 
7.5$\times$ 10$^{21}$ atoms cm$^{-2}$ and the systematics at $0.01$.

Figures 1(a-b) show the variation of the QPO frequencies in (a) the onset and (b) the decline phases
of the outburst. The full widths at half maxima of the fitted QPOs have been used as the error bars.
In the onset phase (a), $0^{th}$ day starts on MJD=53426. The fitted curve represents our solution
(Eqns. 1-2) in which the shock is launched at $r_s=1270$ which drifts slowly at $v_0=1970$cm s$^{-1}$. 
On the $15^{th}$ day after the outburst starts, the noise was high, but we could clearly observe two
different QPO frequencies at a very short time interval. At the time of the last QPO detection 
($15.41^{th}$ day) at $\nu=17.78$Hz, the shock seems to be at $r \approx 59$.
The strength of the shock $R$, which may have began with $R=R_0\sim 4$ is found
to be time dependent, slowly getting weakened on its way to the black hole
($R \rightarrow 1$ as $r\rightarrow r_g$). For simplicity, we assume
$1/R\rightarrow 1/R_0 + \alpha t_d^2$, where $\alpha$ is a very small number limited by the 
time in which the shock disappears (here~$t_{ds} \sim 15.5$days). Thus, $\alpha \sim 1/t_{ds}^2$.
In our solution, $\alpha \sim 0.001$. The fit remains generally very good even with 
a shock of constant strength ($R=R_0$). On the last day of our observation, 
the shock strength went down to $R \sim 2$ at $r\sim 59$.

In the declining phase (Fig. 1b), the QPO frequency on the first day
($MJD=53631$) corresponds to launching the shock at about $r_s=40$. It evolves
as $ \nu_{QPO} \sim t_d^{-0.2}$. Since $ \nu_{QPO} \sim r_s^{-2/3}$ (Eq. 1),
the shock was found to drift very slowly with time ($r_s \sim t_d^{0.13}$) till
about $t_d=3.5$ day where the shock location was $\sim r=59$.
After that it moves out roughly at a constant acceleration ($r_s \sim t_d^{2.3}$)
and the QPO frequency decreases as $\nu_{QPO}\sim r_s^{-2/3} \sim t_d^{-3.5}$.
Finally, when the QPO was last detected, on $t_d=19.92$th day ($MJD=53648$)
the shock went as far as $r_s=3100$ and beyond that it was not detected. 
In Figs. 2(a-b), we present the dynamic PDS where the vertical direction 
indicates the QPO frequency. The intensity in the gray scale signifies power. 
Results of five dwells are given in both the onset and decline phases.
In Fig. 2a, dwell nos. (1-3), and in Fig. 2b dwell nos. (2-5), the gray scale is normalized to
log(Power) = -6 to 0. However, to show more contrast, the gray scale in the remaining dwells
are normalized to log(Power) = -6 to -3.

\begin {figure}
\centering
\begin{tabular}{ccc}
\includegraphics[width=.5\linewidth,angle=-0]{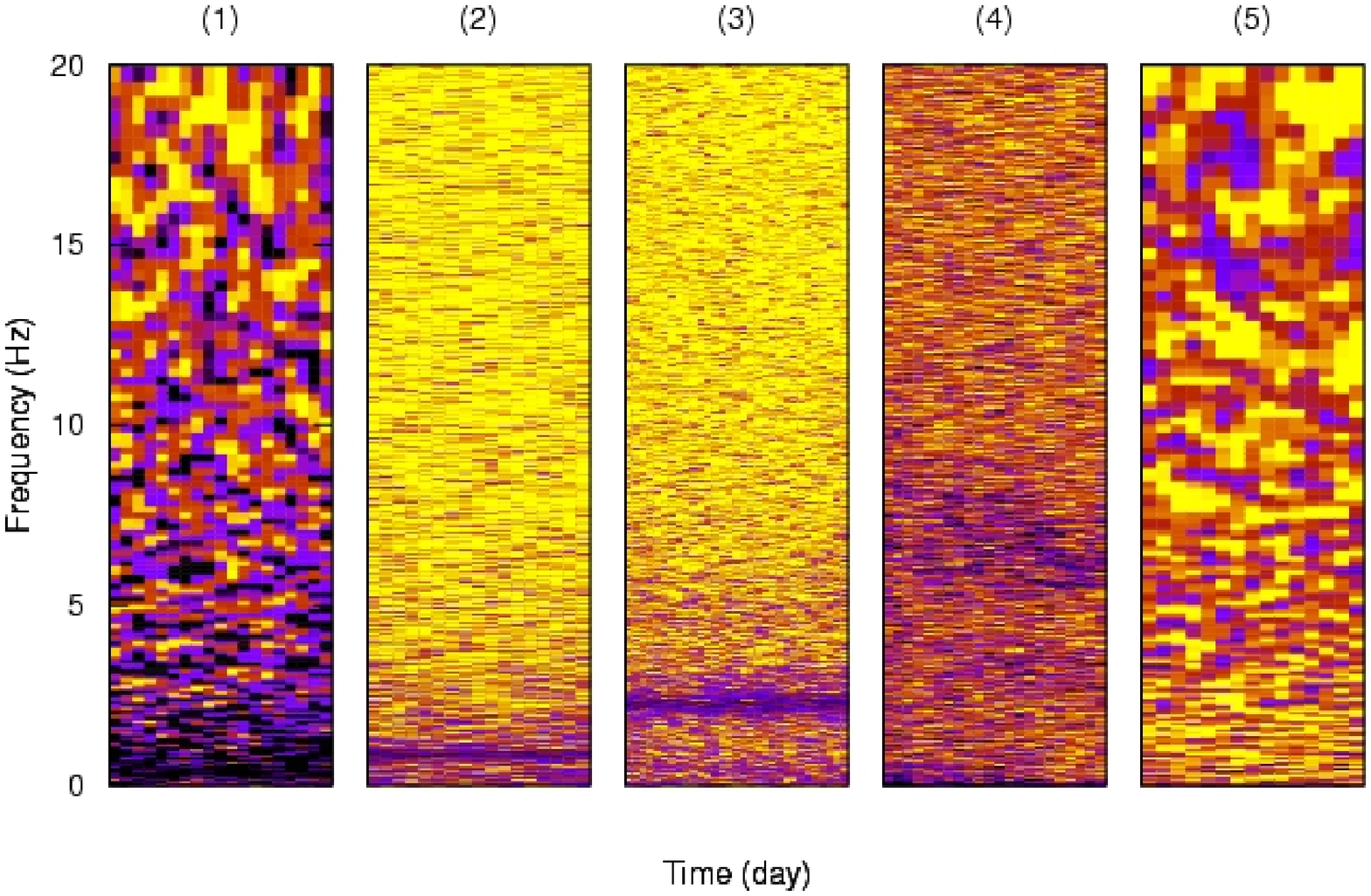} & \hspace{-1cm} &
\includegraphics[width=.5\linewidth,angle=-0]{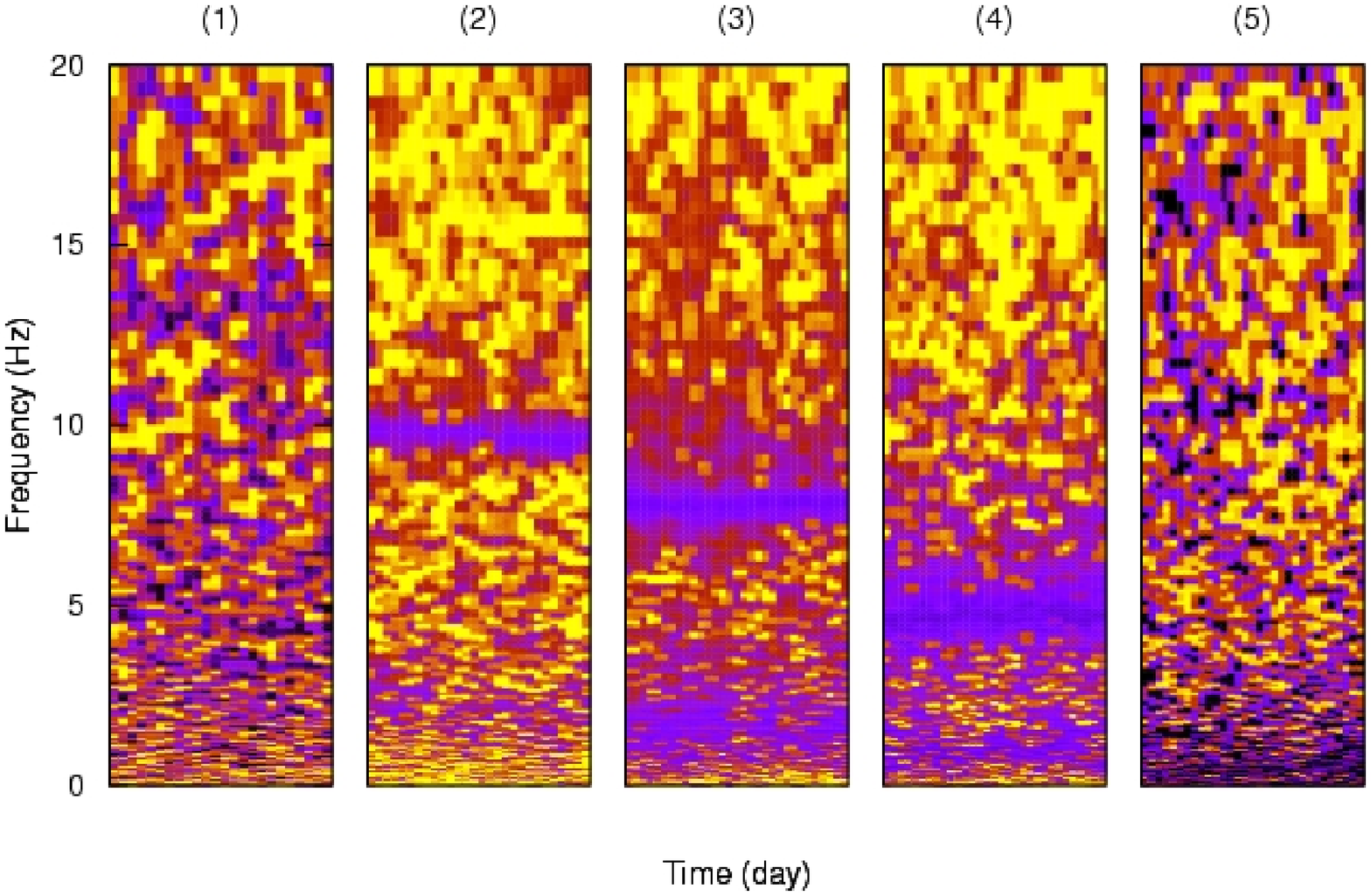}
\end{tabular}
\caption{Fig. 2 (a-b): (a) Dynamic power density spectra in five
days in the onset phase. (1) Obs. ID=91404-01-01-01, MJD=53435.6115, Day=9.57, QPO=0.382 Hz,
(2) Obs. ID=91702-01-01-03, MJD=53438.0539, Day=12.0139, QPO=0.886 Hz,
(3) Obs. ID=90704-04-01-00, MJD=53439.7400, Day=13.7000, QPO=2.3130 Hz,
(4) Obs. ID=91702-01-02-00, MJD=53440.7357, Day=14.6957, QPO=3.45 \& 6.522 Hz with a break frequency
at 0.78 Hz and (5) Obs. ID=91702-01-02-01, MJD=53441.5109, Day=15.4119, QPO=14.54 \& 17.78 Hz.
(b) Dynamic power density spectra in five days in the
decline phase. (1) Obs. ID=91702-01-76-00, MJD=53628.1962, Day=0.00, QPO=13.14 Hz,
(2) Obs. ID=91702-01-79-01, MJD=53629.3760, Day=1.1798, QPO=9.77 Hz,
(3) Obs. ID=91702-01-80-00, MJD=53631.4734, Day=3.2772, QPO=7.823 \& 15.2 Hz with a break
frequency at 1.32 Hz, (4) Obs. ID=91702-01-80-01, MJD=53632.4557, Day=4.2595,
QPO=4.732 Hz with a break frequency=0.86 Hz, (5) Obs. ID=91702-01-82-00,
MJD=53636.4517, Day=8.2555, QPO=0.423 Hz.  }
\end{figure}

We found significant changes in both the timing and the spectral properties
in these two phases. In the rising phase, the spectrum clearly becomes softer 
as the shock moves in and the QPO frequency goes up. In the declining phase, 
as the QPO frequency decreases, all three of the black body (BB), the Comptonization (CST)
and the power-law (PL) components decrease. Ultimately all that remains is basically a
weak power-law component, perhaps due to the faint jet or hot sub-Keplerian flow in the
disk which is left over after the outburst is over. We also find that the disk component
which was becoming stronger at the onset phase is absent in the late declining phase.

\section{Discussions and concluding remarks}

In this article, we analyze the rising and the decline phases of the most recent outburst 
of the black hole candidate GRO J1655-40 and show that during the rising phase, slowly 
drifting shock oscillation solution explains the rise of QPO frequencies very well. In the 
decline phase, the shock propagated outwards.

The solutions we present here is unique in the sense that we are able to connect the 
QPO frequency of one observation with that of the next by a simple analytical means. 
We have shown how waves of matter are disappearing behind the horizon of the black 
hole right after the last day of the rising phase.

\section*{Acknowledgments}

D. Debnath acknowledges the support of a CSIR scholarship and P.S. Pal acknowledges 
the support of an ISRO RESPOND project.

{}

\end{document}